\documentclass[preprint,eqsecnum,aps,showpacs]{revtex4}
\begin{document}
%\input psfig
%\count255=\time\divide\count255 by 60 \xdef\hourmin{\number\count255}
%  \multiply\count255 by-60\advance\count255 by\time
% \xdef\hourmin{\hourmin:\ifnum\count255<10 0\fi\the\count255}
%\draft\preprint{WM-01-110}
\newcommand{\xbf}[1]{\mbox{\boldmath $ #1 $}}
\title{Quadrupole Moments of $N$ and $\Delta$ in the $1/N_c$ Expansion}

\author{Alfons J. Buchmann}
\email{alfons.buchmann@uni-tuebingen.de}
\affiliation{Institute for Theoretical Physics, University of T\"{u}bingen, D-72076, Germany}
\author{Janice A. Hester}
\email{jah1113@netscape.net}
\author{Richard F. Lebed}
\email{Richard.Lebed@asu.edu}
\affiliation{Department of Physics and Astronomy, Arizona State University,
Tempe, AZ 85287-1504}
\date{May, 2002}
\begin{abstract}
We calculate expressions for the quadrupole moments of nonstrange
baryons in which the number of QCD color charges is $N_c$.  Using only
the assumption of single-photon exchange, we obtain 4 relations 
among the 6 moments, and show how all of them may be obtained from
$Q_{\Delta^+ p}$ up to $O(1/N_c^2)$ corrections.  We compare to the
$N_c=3$ case, and obtain relations between the neutron charge radius
and quadrupole moments.  We also discuss prospects for the measurement
of these moments.
\end{abstract}
%11.15.Pg   Expansions for large numbers of components (e.g., 1/Nc expansions)
%12.39.-x   Phenomenological quark models 
%13.40.Em   Electric and magnetic moments 
%14.20.-c   Baryons (including antiparticles)

\maketitle
\thispagestyle{empty}
\newpage
\setcounter{page}{1}

\section{Introduction}\label{sec:intro}
The most interesting observable properties of hadrons, such as masses,
electromagnetic moments, and scattering amplitudes, fall squarely in
the nonperturbative regime of QCD and thus resist first-principles
analytic calculation. While lattice simulations for static observables
continue to improve numerically, they do not typically address
questions regarding which features of QCD explain which various
aspects of hadronic observables.  Other techniques, such as the
operator product expansion, relate perturbative quantities, hadronic
observables, and nonperturbative matrix elements pertaining to vacuum
structure.  The $1/N_c$ expansion~\cite{tHooft}, with $N_c$ the number
of color charges, is alone among QCD-inspired techniques in providing
a perturbative parameter at all energy scales.  While the notion of
varying the fixed value of $N_c = 3$ may seem peculiar on first
glance, the $1/N_c$ expansion is made phenomenologically useful by the
observation that the structure of the physical universe would not be
qualitatively very different were the precise value of 
$N_c$ = 5, 11, or 113~\cite{reviews}.

The $1/N_c$ expansion has proved quite useful in describing the
properties of baryons, such as ground-state masses~\cite{JL}, magnetic
moments~\cite{LMRW,DDJM}, and axial couplings~\cite{DDJM,FJM}, as well
as excited baryon masses~\cite{CCGL,GSS,CC1}, axial
couplings~\cite{CC1,PY,Goity}, and pion-production~\cite{CGKM} and
electromagnetic decay and production~\cite{CC2} amplitudes.  Similar
studies have investigated baryons containing a heavy
quark~\cite{J1,MD}, and the nature of the nucleon-nucleon
interaction~\cite{NN}, as well as a variety of other properties.

The analysis proceeds by studying interaction Hamiltonian operators of
known spin-flavor transformation properties acting upon the baryon
states.  Each such operator is characterized by a coefficient
proportional to a particular power of $1/N_c$. The latter may be
obtained by counting the minimum number of gluon exchanges necessary
to connect the requisite number of quark lines in order for the
operator to have a nonvanishing expectation value for the baryon
state, since $\alpha_s \propto 1/N_c$.  Of course, the possibility
that contributions from a given operator may add coherently over the
$N_c$ quarks, giving an additional enhancement of $O(N_c)$ to the full
matrix element, must be taken into account.  To this scheme one must
add the observation that not all apparently different spin-flavor
operators are linearly independent on a given multiplet of baryons.
Often there appear relations between seemingly distinct operators due
to the structure of the spin-flavor group [SU(6) for three quark
flavors] or due to the fact that one considers operators acting upon a
particular representation of that group (the ground-state {\bf 56} for
$N_c = 3$, for example).  Application of such {\it operator reduction
rules}~\cite{CCGL,DJM} are then necessary to remove redundant degrees
of freedom; see Sec.~\ref{ops} for explicit examples.  Together, these
effects provide a power-counting scheme for deciding which Hamiltonian
operators are most significant in determining contributions to the
physical properties listed above.

While we have been discussing the baryon as an $N_c$-quark state, 
of course the physical baryon is a much more complicated object,
containing a complex soup of gluons and sea quark-antiquark pairs.
The simple-minded picture may be justified by noting that physical
baryons for $N_c=3$ have precisely the same quantum numbers as those
predicted by the simple quark model.  Therefore, one can invent
interpolating fields whose effect is to subsume the entire complicated
substructure into $N_c$ precisely defined constituent quark
fields~\cite{BL1}. This identification allows the old constituent
quark model to be placed on a rigorous footing and elucidates the
nature of the physical picture used in the 
studies noted above.

Once the relevant operators are ordered in a $1/N_c$ expansion and any
operator coefficients due to known suppressions are taken into account
[such as SU(3) flavor or isospin breaking], the matrix elements of the
spin-flavor structures are computed by standard group-theoretical
methods.  Then, only an unknown dimensionless coefficient determined
by strong interaction effects remains uncomputed .  This approach is
none other than the application of the Wigner-Eckart theorem in a
novel context: The matrix elements of the spin-flavor structures
constitute Clebsch-Gordan coefficients, and the leftover coefficients
are reduced matrix elements.  Inasmuch as the expansion is natural,
{\it i.e.}, takes into account all known suppressions or enhancements,
these reduced matrix elements should be of order unity.

In the most general application of the method, one first generates a
complete set of linearly independent operators, which is guaranteed to
exist since the baryons belong to a finite representation for any
particular $N_c$.  Since ultimately one sets $N_c \to 3$, it is
sufficient to consider only up to 3-body operators, {\it i.e.},
irreducible operators involving 3 distinct quark lines.  While this
would seem to offer no predictivity---after all, then one would have
an equal number of observables and operator coefficients---the
hierarchy specified by the $1/N_c$ expansion determines which
operators are more or less phenomenologically significant.

In fact, the method can easily be adapted to various model
assumptions.  Suppose, for example, that one works in a model in which
isospin violation arises from only one source, say, quark charges.
The most general expansion allows for every possible source of isospin
violation; therefore, a number of possible operators can be neglected
in this model.  Since each independent operator is sensitive to a
particular pattern of spin-flavor symmetry breaking that contributes
only to certain combinations of baryon observables, the absence of
each such operator in the model indicates a relation obeyed by the
baryon states.  This is precisely the same as the approach that was
adopted for studying baryon charge radii in Ref.~\cite{BL1}, and for
the $N_c = 3$ case for quadrupole moments in Ref.~\cite {BH2}.

In this work we consider the quadrupole moments of the non-strange
ground-state baryons.  The quadrupole moments hold interest in
providing vital information on the charge distributions of the
baryons, which are inherently non-perturbative quantities.  Only the
non-strange baryons are considered in this work, as explained below,
because the tricks of group theory for the 2-flavor case are different
from those necessary to study the 3-flavor case.  The latter case for
quadrupole moments as well as charge radii will be considered in a
subsequent work~\cite{BL3}.

But are baryon quadrupole moments measurable quantities?  We address
this central issue in Sec.~\ref{meas}.  In Sec.~\ref{ops} we generate
the operator basis for a very general model: The only assumption is
the one-photon exchange approximation, by which we mean that an
arbitrary number of quarks can be involved in the quadrupole operator,
but the photon itself couples to only one of them.  The results of the
analysis are presented in Sec.~\ref{results} and conclusions in
Sec.~\ref{concl}.

\section{Prospects for Measurement} \label{meas}

The dearth of literature on baryon quadrupole moments compared to that
on masses or magnetic moments points to the basic difficulty of
measuring these quantities.  The quadrupole tensor is rank-2, and
therefore by the Wigner-Eckart theorem its expectation value, called
the {\it spectroscopic\/} quadrupole moment $Q$, vanishes for the
spin-1/2 nucleons.  However, it should be noted~\cite{BM} that $Q$,
measured in space-fixed coordinates, is not identical to the second
moment of the charge distribution,
\begin{equation}
e Q_0 \equiv \left< \int d^3 r \, \rho({\bf r}) (3z^2 - r^2) \right> ,
\end{equation}
measured in body-fixed coordinates, with $e$ the proton charge.  $Q_0$
is called the {\it intrinsic\/} quadrupole moment, which would
describe the shape of the charge distribution were it possible to take
a snapshot of the particle at a fixed time.  The most obvious
manifestation of the intrinsic quadrupole moment of a spheroidal
particle is whether it is prolate (football-shaped) or oblate
(pancake-shaped).  The relation between the two definitions for a
particle of ground-state spin $J$ is given by
\cite{BM}:
\begin{equation}
Q = \frac{J(2J-1)}{(J+1)(2J+3)} \, Q_0.
\end{equation}
For $J = 0$ or $1/2$, the particle body-fixed axis points with equal
probability in all directions by parity invariance, leading to the
necessary vanishing of $Q$.  However, $Q_0$ can be nonzero for such
particles, and it has been shown in a variety of models~\cite{BH1}
that the proton is prolate and the $\Delta^+$ is oblate.  It has also
been shown~\cite{Kumar} that, given a sufficient number of measured
electric quadrupole $(E2)$ transition matrix elements, $Q_0$ can be
extracted in a model-independent way. Nevertheless, the quadrupole
moments discussed below are of the spectroscopic variety and are
obtainable through quadrupole transitions.

The spectroscopic quadrupole moments for the ground-state baryons
consist of diagonal transitions for the spin-3/2 decuplet baryons
$\Delta$, $\Sigma^*$, $\Xi^*$, and $\Omega$, and off-diagonal
transitions between the spin-3/2 decuplet and spin-1/2 octet baryons,
$\Delta N$, $\Sigma^* \Sigma$, $\Sigma^* \Lambda$, and $\Xi^* \Xi$.
Since all of the decuplet baryons except the $\Omega$ decay strongly,
couplings of the form decuplet-decuplet-$\gamma$ can be measured only
through virtual processes.  Such experiments are difficult but not
impossible; for example, Ref.~\cite{Drechsel} proposes the measurement
of the $\Delta^+$ magnetic moment through the process $\gamma p
\to\gamma \pi^0 p$.  However, the same authors note the
near-impossibility of measuring the electric quadrupole moment through
such experiments.  The problem is that electric quadrupole ($E2$)
operators are time-reversal odd.  Thus, if the initial state and final
state are identical (as for static diagonal quadrupole transitions),
then this matrix element vanishes.  There are only two ways around
this restriction; the first is to use Coulomb photons rather than real
radiation (the usefulness of Coulomb photons is limited by the short
lifetime of the decuplet baryons), and the second is to extract the
off-shell or recoil effects of the decuplet baryons, a proposition
fraught with many difficulties.  For our purposes, we suppose that the
diagonal quadrupole transitions involving decuplet baryons will not be
measured any time soon, and suggest that our calculation provides
predictions of quantities that are very hard to measure.

The $\Omega^-$ is an exception to this constraint, since it has an
appreciable lifetime.  A number of clever experiments have been
suggested to extract $Q_\Omega$, including the measurement of energy
levels of an exotic atom formed by a heavy nucleus capturing an
$\Omega^-$~\cite{OmAtom} and the precession of $\Omega^-$ spin as it
traverses a crystal at small angles to the crystallographic
plane~\cite{OmCrystal}.  In both cases, the purpose is to enhance the
effect of the electric field gradient, to which the quadrupole moment
couples.

Since all the octet baryons (except the $\Sigma^0$) have appreciable
lifetimes, the transition $E2$ moments are all measurable in
principle, inasmuch as one possesses a suitable source of the desired
long-lived baryon.  In the hyperon sector, the Primakoff reaction $Y +
Z \to Y^* + Z$, where $Y(Y^*)$ is the octet(decuplet) hyperon and $Z$
is a heavy nucleus, is sensitive to both the $M1$ and $E2$ transition
matrix elements; this process has been studied at SELEX~\cite{SELEX}
to obtain a bound on the radiative width of the $\Sigma^*$.
Experiments at Jefferson Lab that involve kaon photoproduction
($\gamma p \to K^+ Y^* \to K^+ Y \gamma$)~\cite{JLab} can also provide
useful information on these amplitudes.

Lastly, the $N \to \Delta$ transition amplitudes have been studied in
numerous experiments, for example see~\cite{LEGS,others}.  In
particular, the transition quadrupole moment $Q_{N \to \Delta}$ may be
extracted from the $E2/M1$ ratio (the relative quadrupole to dipole
strength).  A sample recent measurement using pion photoproduction
data is $[-3.07 \pm 0.26 \, ({\rm stat + syst}) \pm 0.24 \, ({\rm
model})]\%$~\cite{LEGS}, from which one obtains the value $Q_{N\to
\Delta} = -0.108 \pm 0.009 \pm 0.034$ fm$^2$.  However, not all
researchers agree on the size of uncertainties ({\it e.g.}, those
induced by a particular model used to describe the nonresonant
background~\cite{Bartsch}) when extracting the resonance amplitudes
from the measurements.

\section{The Operator Method} \label{ops}

The operator method was described in the Introduction in general
terms; here we lay out the specifics for this calculation.  It is
identical to the approach used for charge radii in \cite{BL1}, and is
the generalization (for nonstrange states) of the calculation
performed in Ref.~\cite{BH2}.  Any spin-flavor operator can be built
from a basis using no more than three 1-body operators, {\it i.e.},
spin-flavor operators acting upon one quark line.  The most general
such operators (restricted, of course, to those that transform as
quadrupole tensors) are formed from sandwiching Pauli spin matrices
and Gell-Mann flavor matrices (or isospin matrices in the 2-flavor
case) between quark creation and destruction operators.  One then
performs the operator reduction; for example, $I^2 = J^2$ on the
2-flavor states.  Were we to compute this model-independent expansion
for the nonstrange baryon quadrupole moments, we would find precisely
6 operators, matching the number observables $Q_{\Delta^{++}}$,
$Q_{\Delta^+}$, $Q_{\Delta^0}$, $Q_{\Delta^-}$, $Q_{\Delta^+ p}$, and
$Q_{\Delta^0 n}$.  Some of the operators in this list are suppressed
by various powers of $1/N_c$, leading to model-independent approximate
relations between the quadrupole moments.  Precisely this sort of
expansion was carried out for nonstrange baryon charge radii
in~\cite{BL1}.

However, we adopt the mildly model-dependent but much more physical
viewpoint that any quadrupole operator involves the quark charge only
once.  Operators containing, {\it e.g.}, extra powers of the quark
charge are suppressed by additional powers of $e^2/4\pi=1/137$
compared to the single-photon exchange approximation. This restricts
the isospin structure of possible quadrupole operators considerably:
Only one very particular combination of isosinglet and isovector
coupling, given by the quark charge operator, then appears.

The quadrupole spin coupling is a rank-2 tensor, and therefore clearly
cannot be built with just one (rank-1) Pauli spin matrix.  The
diagonal  rank-2 combination formed by two Pauli matrices acting on
quarks labeled by $i,j$ is just the familiar form: 
$$ 3 \, \sigma_{iz} \, \sigma_{jz} - \xbf{\sigma}_i \cdot
\xbf{\sigma}_j \ .$$
It is also possible to build a rank-2 spin tensor with 3 Pauli
matrices, but such a form on any given quark pair can be divided into
symmetric and antisymmetric Hermitian parts.  The symmetric part turns
out to be time-reversal odd (since angular momentum is T-odd), and is
thus irrelevant in strong and electromagnetic matrix elements.  The
antisymmetric part is essentially a commutator of spin generators, and
thus reduces to a single Pauli matrix; this is an example of an
operator reduction rule.  Thus, the only spin structure required up to
the 3-body level is the operator with {\it two} Pauli matrices listed
above.

The quark charge operator can now either act upon one of the two quark
lines on which the quadrupole spin tensor acts, or upon a third line.
Using the general rule that the minimum number of gluon exchanges
necessary to connect the quark lines of an $n$-body operator is $n-1$,
the former is a 2-body operator and thus has the suppression
coefficient $1/N_c$, while the latter is a 3-body operator and thus is
prefaced with a $1/N_c^2$.  One finds that only two operators appear
up to the 3-body level:
\begin{equation} \label{quad}
{\cal Q} = \frac{B}{N_c} \sum_{i \neq j}^{N_c} Q_i (3
 \, \sigma_{iz} \, \sigma_{jz} - \xbf{\sigma}_i \cdot\xbf{
\sigma}_j ) + \frac{C}{N_c^2} \sum_{i \neq j \neq k}^{N_c} 
Q_k (3 \, \sigma_{iz} \, \sigma_{jz} - \xbf{\sigma}_i \cdot
\xbf{\sigma}_j ),
\end{equation}
where $B$ and $C$ are unknown coefficients of order unity, times a
characteristic hadronic quadrupole size (in fm$^2$).  This implies in
particular 4 all-orders relations among the 6 quadrupole moments, as
we see below. The coefficient powers of $1/N_c$ are the only explicit
differences between this expression and that in the $N_c = 3$
case~\cite{BH2}.

In deriving expressions for the matrix elements of these two
operators, it is of great advantage to note that the ground-state
baryon representation---the analogue of the {\bf 56} of SU(6)---is
completely symmetric in combined spin and flavor indices.  This means
in particular that all quarks of a given flavor $u$ or $d$ are
completely symmetrized, and hence carry the maximal spin: $S_u= N_u/2$
and $S_d = N_d/2$, where $N_i$ is the number of quarks of flavor $i$
in the baryon. Furthermore, one has the constraints $N_u + N_d = N_c$
and $N_u - N_d = I_3/2$, and for nonstrange states the rule $I=J$
applies, where {\bf J} = {\bf S}$_u$ + {\bf S}$_d$ is the total baryon
spin.  A complete list of compatible operators is thus $J^2 =I^2$,
$J_3$, $I_3$, $S_u{}^2$, and $S_d{}^2$.

In the calculation one encounters operators such as $\sum_i^{N_c}
Q_i\sigma_{iz}/2 = Q_u S_{uz} + Q_d S_{dz}$.  The values of $Q_{u,d}$
are fixed by the anomaly cancellation conditions of the standard model
with gauge group $SU(N_c) \times SU(2) \times U(1)$ to
be~\cite{Shrock}:
\begin{equation}
Q_{u,c,t} = (N_c+1)/2N_c, \ \ Q_{d,s,b} = (-N_c+1)/2N_c .
\end{equation}

The calculation of all the necessary matrix elements for the 2-flavor
case therefore requires only knowledge of two specific matrix elements
for coupled angular momenta $(j_1, j_2) \to J$; of course, here $j_1$
and $j_2$ stand for $S_u$ and $S_d$, the total spin angular momenta
carried by the $u$ and $d$ quarks, respectively.  These matrix
elements are:
\begin{equation}
\left< J M (j_1 \, j_2) \left| J_{1z} \right| J M^\prime (j_1 \, j_2)
\right> = \frac 1 2 M \delta_{M^\prime M} \left[ 1 + \frac{j_1 (j_1+1)
- j_2 (j_2 + 1)}{J(J+1)} \right] ,
\end{equation}
and
\begin{equation}
\left< J M (j_1 \, j_2) \left| J_{1z} \right| J \! - \! 1 \, M^\prime
(j_1 \, j_2) \right> = \frac{\delta_{M^\prime M}}{2 J}
\sqrt{\frac{[J^2- (j_1 - j_2)] [(j_1 + j_2 + 1)^2 - J^2] (J^2 -
M^2)}{(2J+1)(2J-1)}} .
\end{equation}
These expressions can be derived through elementary means.  The first
is a special case of the Wigner-Eckart theorem often called the {\it
projection theorem}~\cite{Sak}.  The second may be obtained solely
through the manipulation of raising and lowering operators and is
proved in, {\it e.g.}, Ref.~\cite{CS}.  Since these theorems refer to
the coupling of just two states of known angular momentum, they
provide a complete answer only in the 2-flavor case.  Of course, the
3-flavor case requires the coupling of an additional angular momentum
({\bf S}$_s$), which introduces $6j$ and $9j$ symbols, as well as the
careful consideration of additional symmetry properties of the baryon
states.  We do not pursue these topics here, but relegate their
resolution to another paper~\cite{BL3}.

The operators {\bf S}$_{u,d}$ are rank-1 tensors and therefore can
connect states differing by up to one unit of total angular momentum
$J$.  Thus one obtains both diagonal and off-diagonal quadrupole
transition matrix elements.  Expressing Eq.~(\ref{quad}) as ${\cal Q}
= B {\cal O}_B + C {\cal O}_C$, one finds:
\begin{eqnarray}
\left< J J \left| {\cal O}_B \right| J J \right> & = &
+ \frac{1}{N_c} \left(J - \frac 1 2 \right) \left[
\frac{(N_c+2)(2Q-1)}{J+1} + \frac{4J}{N_c} \right] , \nonumber \\
\left< J \, J \! - \! 1 \left| {\cal
 O}_B \right| J \! - \! 1 \, J \!-
\! 1 \right> & = &
+ \frac{3(J-1)}{2JN_c}
\sqrt{\frac{[(2Q-1)^2-4J^2][4J^2-(N_c+2)^2]}{2J+1}} , \nonumber \\
\left< J J \left| {\cal O}_C \right| J J \right> & = &
-\frac{2}{N_c} \left( J - \frac 1 2 \right) \left[
\frac{(N_c+2)(2Q-1)}{J+1} - 4J \left( Q - \frac{1}{N_c} \right)
\right] , \nonumber \\
\left< J \, J \! - \! 1 \left| {\cal O}_C 
\right| J \! - \! 1 \, J \!- \! 1 \right> & = &
- \frac{3(J-1)}{JN_c}\sqrt{\frac{[(2Q-1)^2-4J^2]
[4J^2-(N_c+2)^2]}{2J+1}} , \label{eval}
\end{eqnarray}
where $Q$ is the total baryon electric charge.

\section{Results and Discussion} \label{results}

Equations~(\ref{eval}) evaluated for the quantum numbers of the 6
nonstrange states give the results listed in Table~I.  First note that
these results agree with those in the $N_c = 3$ case, Tables~I and II
of Ref.~\cite{BH2}, except for our aforementioned coefficient factors
of $1/N_c$ and $1/N_c^2$ for $B$ and $C$, respectively, and a factor
of 2 accidentally neglected in the $C$ terms of~\cite{BH2}.  The
matrix elements in \cite{BH2} were obtained by using explicit $N_c=3$
spin-flavor baryon wave functions.

Several features are immediately obvious.  The diagonal matrix
elements for $N_c = 3$ are just $4Q(3B+C)/9$, while those in the
$N_c\to \infty$ limit are $4I_3B/5 = 4(Q-1/2)B/5$, and the two
off-diagonal elements are equal.  These simple linear patterns arise
from our model assumption that only one quark charge operator appears,
and so the quadrupole operator transforms only as $I=0$ and 1.  In
particular, the $I=3/2$ $\Delta$'s can be connected by operators
transforming as $I=0,1,2,$ or 3, and the $I=2$ and 3 combinations must
vanish for any $N_c$.  These combinations are:
\begin{eqnarray}
Q_{\Delta^{++}} - 3 Q_{\Delta^+} + 3 Q_{\Delta^0} - Q_{\Delta^-} & =
&0 \ \ (I=3) , \nonumber \\ Q_{\Delta^{++}} - Q_{\Delta^+} -
Q_{\Delta^0} + Q_{\Delta^-} & = & 0 \ \ (I=2) .
\end{eqnarray}
Quadrupole transition operators between the $I=3/2$ $\Delta$ and
$I=1/2$ $N$ can transform as $I=1$ or 2, and since the latter does not
appear in our model, it leads to the relation:
\begin{equation}
Q_{\Delta^+ p} = Q_{\Delta^0 n} ,
\end{equation}
for all $N_c$.  The last of the 4 relations (6 degrees of freedom, 2
operators) is $N_c$ dependent:
\begin{equation}
\frac{Q_{\Delta^0}}{Q_{\Delta^{+}p}} = -\frac{2}{5} \frac{N_c-3}{N_c}
\sqrt{\frac{2(N_c+5)}{N_c-1}} ,
\end{equation}
and in particular vanishes 
for $N_c = 3$.

In addition to combinations that vanish identically, one may consider
combinations for which the leading in $1/N_c$ terms cancel.  In
particular, we note that generically, the quadrupole moments for large
$N_c$ are $O(N_c^0)$, the sole leading contribution arising from the
operator ${\cal O}_B$.  As discussed above, the diagonal matrix
elements will prove extremely difficult to measure directly, and so we
now seek to express all other quadrupole moments in terms of the one
most easily measurable, $Q_{\Delta^+ p}$.  In terms of the combination
${\cal Q} = 2\sqrt{2} Q_{\Delta^+ p}/5$, we find:
\begin{eqnarray}
Q_{\Delta^{++}} & = & +3 {\cal Q} \left[ 1 + \frac{1} {N_c^2}
\left(\frac{19}{2} + 20 \frac C B \right) + O\left( \frac{1}{N_c^3}
\right) \right] ,
\nonumber \\
Q_{\Delta^+} & = & +{\cal Q} \left[ 1 + \frac{1}{N_c^2} \left(
\frac{39}{2} + 30 \frac C B \right) + O\left( \frac{1}{N_c^3}
\right) \right] , \nonumber \\
Q_{\Delta^0} & = & -{\cal Q} \left[ 1 - \frac{21}{2N_c^2} +
O\left( \frac{1}{N_c^3} \right) \right] , \nonumber \\
Q_{\Delta^-} & = & -3{\cal Q} \left[ 1- \frac{1}{N_c^2} 
\left( \frac 1 2 - 10 \frac C B \right) 
+ O\left( \frac{1}{N_c^3} \right) \right] . \label{pred}
\end{eqnarray}
Phenomenological experience with the $1/N_c$ expansion tells us that
quantities for which corrections are only $O(1/N_c^2)$ tend to agree
well with their predicted central values.  We should note, however,
that the numerical coefficients of the $O(1/N_c^2)$ terms in this case
can be large for $N_c = 3$, depending upon the precise value of the
ratio $C/B$.  An interesting difference of prediction between this
work and \cite{BH2} is the ratio $Q_{\Delta^+} / Q_{\Delta^+ p}$,
which here is $2\sqrt{2}/5 + O(1/N_c^2)$, and in Eq.~(5) of \cite{BH2}
is $\sqrt{2}$; the latter prediction is obtained by setting $C=0$ (and
of course $N_c=3$).  We defer detailed numerical analysis to such time
as the 3-flavor predictions are also in hand~\cite{BL3}.

Lastly, suppose that the baryon charge radii, which come from
spin-spin terms in the baryon Hamiltonian, arise from the same source
as the quadrupole operators; then the matrix elements for the two
observables appear in fixed ratios.  This occurs, for example, in
one-gluon exchange picture, in which a multipole expansion of the
baryon charge density operator $\rho$ reads:
\begin{eqnarray}
\rho & = & A \sum_i^{N_c} Q_i -\frac{B}{N_c}\sum_{i \neq j}^{N_c}
Q_i \left[  2\xbf{\sigma}_i \cdot \xbf{\sigma}_j
-( 3 \xbf{\sigma}_{i\, z} \, \xbf{\sigma}_{j \, z} 
-\xbf{\sigma}_i \cdot \xbf{\sigma}_j ) \right] \nonumber \\ 
&& -\frac{C}{N_c^2}\sum_{i \neq j \neq k}^{N_c}
Q_k \left[  2\xbf{\sigma}_i \cdot \xbf{\sigma}_j
-( 3 \xbf{\sigma}_{i\, z} \, \xbf{\sigma}_{j \, z} 
-\xbf{\sigma}_i \cdot \xbf{\sigma}_j ) \right].
\end{eqnarray}
Note that the normalization of the spin-spin operator is just $-2$
times that used in Ref.~\cite{BL1}.  Thus, the results obtained there
can be carried directly over for comparison.  In particular, the
neutron charge radius is:
\begin{equation}
r_n^2 = \left( B - 2 
\frac{C}{N_c} \right)
\frac{(N_c-1)(N_c+3)}{N_c^2} ,
\end{equation}
from which we immediately see that
\begin{equation} \label{AJBold}
Q_{\Delta^+ p} = \frac{1}{\sqrt{2}} \, r_n^2 \; \frac{N_c}{N_c+3}
\sqrt{\frac{N_c+5}{N_c-1}} .
\end{equation}
The factor on the r.h.s.\ is especially interesting since it equals 1
in both the $N_c = 3$ and $N_c \to \infty$ cases (and in between never
differs from unity by more than 1.2\%).  Thus, one expects that this
relation should hold especially well.  The $N_c=3$ version was first
derived in Ref.~\cite{BHF} using a constituent quark model.  Using the
value $r_n^2 = -0.113(3)(4)$ fm$^2$~\cite{rn2}, one predicts
$Q_{\Delta^+ p} = -0.0799(4)$ fm$^2$, which agrees well with the value
stated in Sec.~\ref{meas}~\cite{LEGS}.  One could continue from here
to predict all the baryon quadrupole moments using
Eq.~(\ref{pred}). The reader should be reminded, however, that the
assumption of one-gluon exchange is much more particular than the
minimal assumption used to obtain Table~I.

\section{Conclusions} \label{concl}

The $1/N_c$ expansion provides an additional handle on nonperturbative
QCD phenomenology, through the observation that universes with odd
$N_c > 3$ are very similar to our own.  Using a rigorous definition of
constituent quarks obtainable in the large-$N_c$ limit and a minimal
model ansatz---that quadrupole operators are proportional to the quark
charge (single-photon exchange approximation)---we have shown that the
6 quadrupole moments of the nonstrange baryons can be described by
just 2 distinct operators, leading to a number of relations that hold
for all $N_c$ and others that hold up to $O(1/N_c^2)$ corrections.  We
have also seen that in a one-gluon exchange picture, all of these can
be predicted using the measured value of the neutron charge radius
$r_n^2$.  Detailed analysis of charge radii and quadrupole moments for
the 3-flavor case is forthcoming.

\section*{Acknowledgments}

A.J.B.\ thanks the Deutsche Forschungsgemeinschaft for some support
under title BU 813/3-1.  R.F.L.\ thanks the U.S.\ Department of Energy
for support under Grant No.\ DE-AC05-84ER40150.

\clearpage

\begin{table}
\caption{Quadrupole moments of the nonstrange ground-state baryons for
arbitrary $N_c$, $N_c = 3$, and $N_c \to \infty$.}
\begin{tabular}{l|c|c|c}
\hline\hline
$Q_{\Delta^{++}}$ & $\frac{B}{N_c^2} \left( +\frac 6 5 \right) (N_c^2 +
2N_c + 5) + \frac{C}{N_c^3} \left( -\frac{12}{5} \right) (N_c^2 - 8
N_c + 5)$ & $\frac 8 3 B + \frac 8 9 C$ & $+\frac 6 5 B$ \\
$Q_{\Delta^+}$ & $\frac{B}{N_c^2} \left( +\frac 2 5 \right) (N_c^2 +
2N_c + 15) + \frac{C}{N_c^3} \left( -\frac{4}{5} \right) (N_c^2 -
13N_c + 5)$ & $\frac 4 3 B + \frac 4 9 C$ & $+\frac 2 5 B$ \\
$Q_{\Delta^0}$ & $\frac{B}{N_c^2} \left( -\frac 2 5 \right)
(N_c+5)(N_c-3) + \frac{C}{N_c^3} \left( +\frac{4}{5} \right) (N_c +5)
(N_c-3)$ & 0 & $-\frac 2 5 B$ \\
$Q_{\Delta^-}$ & $\frac{B}{N_c^2} \left( -\frac 6 5 \right)
(N_c^2+2N_c - 5) + \frac{C}{N_c^3} \left( + \frac{12}{5} \right)
(N_c^2-3N_c-5)$ & $-\frac 4 3 B -\frac 4 9 C$ & $-\frac 6 5 B$
 \\
\hline
$Q_{\Delta^+ p}$ & $\left( \frac{B}{N_c} - 2 \frac{C}{N_c^2} \right)
\sqrt{\frac{(N_c+5)(N_c-1)}{2}}$ & $\frac{2\sqrt{2}}{9} (3B-2C)$ &
$\frac{1}{\sqrt{2}} B$ \\
$Q_{\Delta^0 n}$ & $\left( \frac{B}{N_c} - 2 \frac{C}{N_c^2} \right)
\sqrt{\frac{(N_c+5)(N_c-1)}{2}}$ & $\frac{2\sqrt{2}}{9} (3B-2C)$ &
$\frac{1}{\sqrt{2}} B$ \\
\hline\hline
\end{tabular}
\end{table}

\end{document}